\documentclass[twocolumn,tightenlines,superscriptaddress,showpacs,byrevtex]{revtex4}
\usepackage{graphicx}
\usepackage{epsfig}
\usepackage{dcolumn}
\usepackage{bm}
\usepackage{color}

\newcommand{\lum}{{\cal L}}

\newcommand{\BR}{{\cal B}}

\newcommand{\EE}{e^+e^-}

\newcommand{\penz}{\mathrm{\Theta}(1540)^0}
\newcommand{\penp}{\mathrm{\Theta}(1540)^+}
\newcommand{\penpp}{\mathrm{\Theta}(1540)^{++}}

\newcommand{\beq}{\begin{equation}}
\newcommand{\eeq}{\end{equation}}
\newcommand{\bitm}{\begin{itemize}}
\newcommand{\eitm}{\end{itemize}}

\def\Journal#1#2#3#4{{#1} {\bf #2}, #3 (#4)}

\def\EPJC{Eur. Phys. J. C}

\begin{document}
\title{\quad\\[1.0cm]
 First observation of
 $\gamma \gamma \to p \bar{p} K^+ K^-$ and search for exotic baryons in $pK$ systems}


\noaffiliation
\affiliation{University of the Basque Country UPV/EHU, 48080 Bilbao}
\affiliation{Beihang University, Beijing 100191}
\affiliation{University of Bonn, 53115 Bonn}
\affiliation{Budker Institute of Nuclear Physics SB RAS, Novosibirsk 630090}
\affiliation{Faculty of Mathematics and Physics, Charles University, 121 16 Prague}
\affiliation{Chonnam National University, Kwangju 660-701}
\affiliation{University of Cincinnati, Cincinnati, Ohio 45221}
\affiliation{Deutsches Elektronen--Synchrotron, 22607 Hamburg}
\affiliation{University of Florida, Gainesville, Florida 32611}
\affiliation{Justus-Liebig-Universit\"at Gie\ss{}en, 35392 Gie\ss{}en}
\affiliation{Gifu University, Gifu 501-1193}
\affiliation{SOKENDAI (The Graduate University for Advanced Studies), Hayama 240-0193}
\affiliation{Gyeongsang National University, Chinju 660-701}
\affiliation{Hanyang University, Seoul 133-791}
\affiliation{University of Hawaii, Honolulu, Hawaii 96822}
\affiliation{High Energy Accelerator Research Organization (KEK), Tsukuba 305-0801}
\affiliation{IKERBASQUE, Basque Foundation for Science, 48013 Bilbao}
\affiliation{Indian Institute of Science Education and Research Mohali, SAS Nagar, 140306}
\affiliation{Indian Institute of Technology Bhubaneswar, Satya Nagar 751007}
\affiliation{Indian Institute of Technology Guwahati, Assam 781039}
\affiliation{Indian Institute of Technology Madras, Chennai 600036}
\affiliation{Indiana University, Bloomington, Indiana 47408}
\affiliation{Institute of High Energy Physics, Chinese Academy of Sciences, Beijing 100049}
\affiliation{Institute of High Energy Physics, Vienna 1050}
\affiliation{Institute for High Energy Physics, Protvino 142281}
\affiliation{INFN - Sezione di Torino, 10125 Torino}
\affiliation{J. Stefan Institute, 1000 Ljubljana}
\affiliation{Kanagawa University, Yokohama 221-8686}
\affiliation{Institut f\"ur Experimentelle Kernphysik, Karlsruher Institut f\"ur Technologie, 76131 Karlsruhe}
\affiliation{King Abdulaziz City for Science and Technology, Riyadh 11442}
\affiliation{Korea Institute of Science and Technology Information, Daejeon 305-806}
\affiliation{Korea University, Seoul 136-713}
\affiliation{Kyungpook National University, Daegu 702-701}
\affiliation{\'Ecole Polytechnique F\'ed\'erale de Lausanne (EPFL), Lausanne 1015}
\affiliation{P.N. Lebedev Physical Institute of the Russian Academy of Sciences, Moscow 119991}
\affiliation{Faculty of Mathematics and Physics, University of Ljubljana, 1000 Ljubljana}
\affiliation{Ludwig Maximilians University, 80539 Munich}
\affiliation{University of Maribor, 2000 Maribor}
\affiliation{Max-Planck-Institut f\"ur Physik, 80805 M\"unchen}
\affiliation{School of Physics, University of Melbourne, Victoria 3010}
\affiliation{University of Miyazaki, Miyazaki 889-2192}
\affiliation{Moscow Physical Engineering Institute, Moscow 115409}
\affiliation{Moscow Institute of Physics and Technology, Moscow Region 141700}
\affiliation{Graduate School of Science, Nagoya University, Nagoya 464-8602}
\affiliation{Kobayashi-Maskawa Institute, Nagoya University, Nagoya 464-8602}
\affiliation{Nara Women's University, Nara 630-8506}
\affiliation{National Central University, Chung-li 32054}
\affiliation{National United University, Miao Li 36003}
\affiliation{Department of Physics, National Taiwan University, Taipei 10617}
\affiliation{H. Niewodniczanski Institute of Nuclear Physics, Krakow 31-342}
\affiliation{Niigata University, Niigata 950-2181}
\affiliation{University of Nova Gorica, 5000 Nova Gorica}
\affiliation{Novosibirsk State University, Novosibirsk 630090}
\affiliation{Osaka City University, Osaka 558-8585}
\affiliation{Pacific Northwest National Laboratory, Richland, Washington 99352}
\affiliation{University of Pittsburgh, Pittsburgh, Pennsylvania 15260}
\affiliation{University of Science and Technology of China, Hefei 230026}
\affiliation{Seoul National University, Seoul 151-742}
\affiliation{Showa Pharmaceutical University, Tokyo 194-8543}
\affiliation{Soongsil University, Seoul 156-743}
\affiliation{Sungkyunkwan University, Suwon 440-746}
\affiliation{School of Physics, University of Sydney, New South Wales 2006}
\affiliation{Department of Physics, Faculty of Science, University of Tabuk, Tabuk 71451}
\affiliation{Tata Institute of Fundamental Research, Mumbai 400005}
\affiliation{Excellence Cluster Universe, Technische Universit\"at M\"unchen, 85748 Garching}
\affiliation{Department of Physics, Technische Universit\"at M\"unchen, 85748 Garching}
\affiliation{Toho University, Funabashi 274-8510}
\affiliation{Department of Physics, Tohoku University, Sendai 980-8578}
\affiliation{Earthquake Research Institute, University of Tokyo, Tokyo 113-0032}
\affiliation{Department of Physics, University of Tokyo, Tokyo 113-0033}
\affiliation{Tokyo Institute of Technology, Tokyo 152-8550}
\affiliation{Tokyo Metropolitan University, Tokyo 192-0397}
\affiliation{University of Torino, 10124 Torino}
\affiliation{Utkal University, Bhubaneswar 751004}
\affiliation{Virginia Polytechnic Institute and State University, Blacksburg, Virginia 24061}
\affiliation{Wayne State University, Detroit, Michigan 48202}
\affiliation{Yamagata University, Yamagata 990-8560}
\affiliation{Yonsei University, Seoul 120-749}
  \author{C.~P.~Shen}\affiliation{Beihang University, Beijing 100191} 
  \author{C.~Z.~Yuan}\affiliation{Institute of High Energy Physics, Chinese Academy of Sciences, Beijing 100049} 
  \author{I.~Adachi}\affiliation{High Energy Accelerator Research Organization (KEK), Tsukuba 305-0801}\affiliation{SOKENDAI (The Graduate University for Advanced Studies), Hayama 240-0193} 
  \author{H.~Aihara}\affiliation{Department of Physics, University of Tokyo, Tokyo 113-0033} 
  \author{D.~M.~Asner}\affiliation{Pacific Northwest National Laboratory, Richland, Washington 99352} 
  \author{V.~Aulchenko}\affiliation{Budker Institute of Nuclear Physics SB RAS, Novosibirsk 630090}\affiliation{Novosibirsk State University, Novosibirsk 630090} 
  \author{T.~Aushev}\affiliation{Moscow Institute of Physics and Technology, Moscow Region 141700} 
  \author{R.~Ayad}\affiliation{Department of Physics, Faculty of Science, University of Tabuk, Tabuk 71451} 
  \author{V.~Babu}\affiliation{Tata Institute of Fundamental Research, Mumbai 400005} 
  \author{I.~Badhrees}\affiliation{Department of Physics, Faculty of Science, University of Tabuk, Tabuk 71451}\affiliation{King Abdulaziz City for Science and Technology, Riyadh 11442} 
% ...
\collaboration{The Belle Collaboration}

\begin{abstract}

The process $\gamma \gamma \to p \bar{p} K^+ K^-$ and its intermediate processes are measured
for the first time using a 980~fb$^{-1}$ data sample collected with the
Belle detector at the KEKB asymmetric-energy $e^+e^-$ collider.
The production of $p \bar{p} K^+ K^-$ and a $\Lambda(1520)^0~(\bar{\Lambda}(1520)^0)$ signal in the
$pK^-$~($\bar{p} K^+$) invariant mass spectrum are clearly observed.
However, no evidence for an exotic baryon near 1540~MeV/$c^2$,
denoted as $\penz$~($\bar{\Theta}~(1540)^0$) or $\penpp$~($\Theta(1540)^{--}$),
is seen in the $p K^-$~($\bar{p}K^+$) or $pK^+$~($\bar{p} K^-$) invariant mass spectra.
Cross sections for $\gamma \gamma \to p \bar{p} K^+ K^-$, $\Lambda(1520)^0 \bar{p} K^+ +c.c.$
and the products $\sigma(\gamma \gamma \to \penz \bar{p} K^+ +c.c.)\BR(\penz\to p K^{-})$ and
$\sigma(\gamma \gamma \to \penpp \bar{p} K^- +c.c.)\BR(\penpp\to p K^{+})$ are measured.
We also determine upper limits on the products of the $\chi_{c0}$ and $\chi_{c2}$
two-photon decay widths and their branching fractions to
$p \bar{p} K^+ K^-$ at the  90\% credibility level.

\end{abstract}


\maketitle


Quantum chromodynamics allows for the existence of exotic hadronic states
such as glueballs, hybrids and multi-quark states with valence quark
and gluon configurations that are distinct from normal
quark-antiquark mesons and three-quark baryons~\cite{gell}. There is a
long history of searches for these types of states, but
no solid examples were seen prior to the recent discovery of
tetraquark and pentaquark states containing charm and beauty quarks.

A charged charmoniumlike state $Z(4430)^{\pm}$ was observed
by the Belle experiment in 2007 in the $\pi^{\pm} \psi'$
system produced in $B$ decays~\cite{zc4430}.
In addition to the $Z(4430)^{\pm}$,
recently BESIII and Belle also observed a series of charged $Z$ states
such as the $Z(3900)^{\pm}$~\cite{belley_new}, $Z(4020)^{\pm}$~\cite{zc4020}, $Z(4200)^{\pm}$~\cite{zc4200},
$Z(4050)^{\pm}$ and $Z(4250)^{\pm}$~\cite{roman}.
As such, a charged state must contain at least four quarks; these $Z$ states have been interpreted either as
tetraquark states, molecular states,
or other configurations~\cite{QWG}.

Very recently, the LHCb collaboration reported the observation of two
exotic structures, denoted as $P_c(4380)^+$ and $P_c(4450)^+$, in
the $J/\psi p$ system in $\Lambda_b^0 \to J/\psi K^-
p$~\cite{lhcb}. Since the valence structure of $J/\psi p$ is $c\bar{c}uud$, the newly discovered
particles must consist of at least five quarks.
Several theoretical interpretations of these states have been
developed, such as the diquark picture~\cite{doublet} and hadronic molecules~\cite{mole}.
Actually, the first strong experimental evidence for a pentaquark state,
referred to as the $\mathrm{\Theta}(1540)^+$, was reported in the
reaction $\gamma n \to n K^+ K^-$ in the LEPS
experiment~\cite{leps}.
It was a candidate for a $uudd\bar{s}$ pentaquark state.
However, it was not confirmed in larger-statistics data samples in the same experiment
and was most probably not a genuine
state~\cite{mao}.

To confirm the pentaquark states discovered by LHCb, further experimental searches for exotic baryons
should be pursued. If fully confirmed, exotic baryons would be most naturally explained as pentaquark states.
The possibility of  observing additional hypothetical exotic baryons in $\gamma \gamma$
collisions is discussed in Ref.~\cite{wu}, where Fig.~3 depicts possible diagrams for
the exclusive double- and single- pentaquark productions in $\gamma \gamma$ collisions.
High luminosity electron-positron colliders are well suited to measurements of the
two-photon production since they provide a large flux of quasi-real photons colliding
at two-photon center-of-mass (CM) energies covering a wide range. Due to the high luminosity
accumulated at {\it B} factories, searches for exotic baryons
in exclusive $\gamma \gamma$ reactions are possible.
The authors in Ref.~\cite{wu} suggest that
single pentaquark production may be viewed as a collision
of a non-resonant di-baryon and a di-meson pair.
One of the
incoming photons fluctuates into two mesons with strangeness; one meson then
collides with the hadronic system produced by the
other incoming photon.
The cross section for the reaction $\gamma \gamma \to p \bar{p} K^+ K^-$
is predicted to be around 0.1 nb for $W_{\gamma \gamma}\ge 2(m_p+m_K)$~\cite{wu}, where
$W_{\gamma \gamma}$ is the CM energy of two-photon system that is the same
as the invariant mass of the final-state hadron system
and $m_p$ and $m_K$
are the proton and kaon nominal masses~\cite{PDG}.
This presents the opportunity to search for novel
exotic baryons, denoted as $\penz\to p K^-$
and $\penpp \to p K^+$ which are similar to $\penp$,  in intermediate processes in two-photon annihilations.

Here, we report the cross sections for
$\gamma\gamma\to p \bar{p} K^+ K^-$ through the measurement of
$e^+e^- \to (e^+e^-) p \bar{p} K^+ K^-$, as well as
searches for possible exotic baryons as intermediate states.
The results are based on an analysis of a 980~fb$^{-1}$ data sample taken at or
near the $\Upsilon(nS)$ ($n=1,...,5$) resonances with the Belle detector~\cite{Belle}
operating at the KEKB asymmetric-energy $\EE$ collider~\cite{KEKB}.
The analysis is made in the ``zero-tag" mode, where neither the recoil
electron nor positron is detected. We restrict the virtuality of the
incident photons to be small by imposing a strict transverse-momentum
balance along the beam axis for the final state hadronic system.

The detector, which is described in detail elsewhere~\cite{Belle}, is a
large-solid-angle magnetic spectrometer that consists of a silicon
vertex detector, a 50-layer central drift chamber, an
array of aerogel threshold Cherenkov counters, a barrel-like
arrangement of time-of-flight scintillation counters, and an
electromagnetic calorimeter comprised of CsI(Tl) crystals
located inside a superconducting solenoid coil that provides a 1.5~T
magnetic field. An iron flux-return located outside of the coil is
instrumented to detect $K_L^0$ mesons and to identify muons.

We use the program {\sc treps}~\cite{treps} to generate signal
Monte Carlo (MC) events and determine experimental efficiencies
and effective luminosities for photon-photon collisions.
In this generator, the two-photon luminosity
function is calculated and events are generated at a
specified $\gamma \gamma$ CM energy ($W_{\gamma \gamma}$)
using the equivalent photon approximation~\cite{berger}. The
efficiencies for detecting $\gamma \gamma \to p \bar{p} K^+ K^-$
and its intermediate processes $\Lambda(1520)^0 \bar{p} K^+$,
$\penz \bar{p} K^+$  and $\penpp \bar{p} K^-$~\cite{charge} at
different fixed $W_{\gamma \gamma}$ values
are determined by assuming a phase space model.
The width of the $\Theta(1540)$ for the nominal results is set to
9.7 MeV from Ref.~\cite{mao}.



We require four reconstructed charged tracks with zero net charge.
For these tracks, the impact parameters perpendicular to and along
the beam direction with respect to the interaction point are
required to be less than 0.5~cm and 4~cm, respectively, and the
transverse momentum in the laboratory frame is restricted to be
higher than 0.1~GeV/$c$. For each charged track, information from
different detector subsystems is combined to form a likelihood
$\mathcal{L}_i$ for each particle species~\cite{pid}.
Tracks of interest with $\mathcal{R}_K=\mathcal{L}_K/(\mathcal{L}_K+\mathcal{L}_\pi)>0.6$ are
identified as kaons with an efficiency of about 91\%;
about 5.9\% of the pions are misidentified as kaons.
A track with $\mathcal{R}_{p/\bar{p}} =
\mathcal{L}_{p/\bar{p}}/(\mathcal{L}_{p/\bar{p}}+\mathcal{L}_\pi) > 0.6$
and $\mathcal{R}_{p/\bar{p}} =
\mathcal{L}_{p/\bar{p}}/(\mathcal{L}_{p/\bar{p}}+\mathcal{L}_K) > 0.6$
is identified as a proton/anti-proton with an efficiency of about 95\%.
A similar likelihood ratio is formed for electron identification~\cite{EID}.
Photon conversion backgrounds are removed
by vetoing any charged track in the event that is
identified as electron or positron ($\mathcal{R}_e >
0.9$); this requirement keeps
more than 98.5\% of the signals.

There are some obvious $e^+e^- \to p \bar{p} K^+ K^-$  backgrounds
from initial state radiation processes
observed as a clear peak at around zero in
the distribution of the square of the missing mass recoiling against
 the four charged tracks $p \bar{p} K^+ K^-$ ($M^2_{\rm miss}$).
We require $M^2_{\rm miss}>4$ (GeV/$c^2$)$^2$ to remove such backgrounds.
To suppress backgrounds with extra neutral clusters
in the final state, events are removed if there are one or more additional
photons with energy greater than 150 MeV.

\begin{figure}[htbp]
\includegraphics[height=8cm,angle=-90]{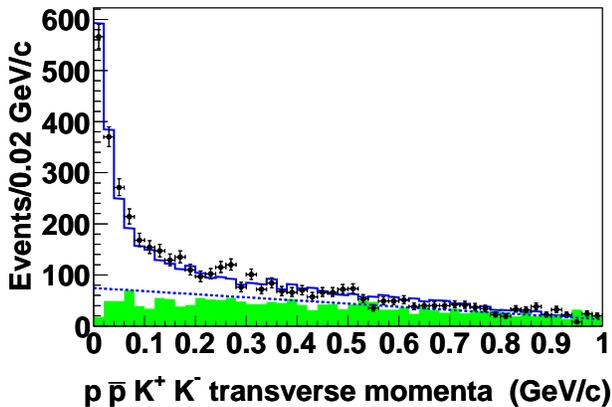}
\caption{\label{xpt-fit}
The magnitude of the
transverse momentum of the $p \bar{p} K^+ K^-$ system with respect to the
beam direction in the $\EE$ CM frame for the selected $p \bar{p} K^+ K^-$
events over the entire mass region. Points with error bars are data. The
solid histogram is the fitted result, the dashed line is the total
background estimate and the shaded histogram is the
normalized contribution from inclusive MC samples described in the
text. }
\end{figure}

The magnitude of the vector sum
of the four transverse momenta in the $\EE$ CM frame, $|\sum {\vec
P}_t^{\ast}|$, which approximates the transverse momentum of the
two-photon-collision system, is used as a discriminating variable
to separate signal from background. The signal tends to
accumulate at small $|\sum {\vec P}_t^{\ast}|$ values while the
non-$\gamma \gamma$ background is distributed over a wider range.
The $|\sum {\vec P}_t^{\ast}|$ distribution for the whole
$p \bar{p} K^+ K^-$ mass region
for the selected events
is shown in Fig.~\ref{xpt-fit}.
This distribution is fitted with a signal shape determined from MC simulation
and a first-order Chebyshev polynomial
to represent the background from non-$\gamma\gamma$ processes.
To check possible peaking backgrounds,
inclusive MC samples of $e^+e^- \to q \bar{q}$ ($q=u,d,c,s$),
$\tau^+ \tau^-$ and $B\bar{B}$ are analyzed.
The shaded histogram in Fig.~\ref{xpt-fit} shows the normalized background contribution
after all selection criteria are applied, which is
consistent with the total background estimated from the fit.

\begin{figure}[htbp]
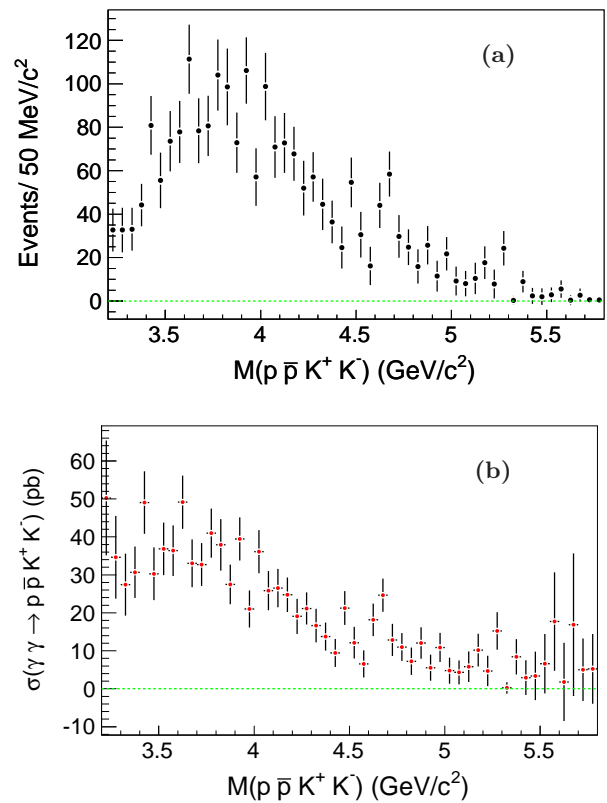

 \psfig{file=fig/m2p2k-fit.epsi,width=5cm, angle=-90}
  \put(-50,-20){ \bf (a)}\vspace{0.5cm}
 \psfig{file=fig/cs-ppkk.epsi,width=5cm, angle=-90}
 \put(-50,-20){ \bf (b)}
\caption{(a) The $p \bar{p} K^+ K^-$ invariant mass distribution obtained by fitting the $|\sum
{\vec P}_t^{\ast}|$  distributions in each $p \bar{p} K^+ K^-$ mass
bin and (b) the cross section of $\gamma \gamma
 \to p \bar{p} K^+ K^-$ in each $p \bar{p} K^+ K^-$ mass bin. The error bars are
statistical only. } \label{m2p2k-fit}
\end{figure}

We obtain the number of $p \bar{p} K^+ K^-$ events in each $p \bar{p} K^+ K^-$ invariant mass
bin  ($n^{\rm fit}$) by fitting the $|\sum {\vec P}_t^{\ast}|$ distribution between zero and 1.0 GeV/$c$.
The signal shape is obtained from MC simulation in the corresponding
mass bin and the background shape is parameterized as a first-order Chebyshev
polynomial. The background shape is fixed to that from the overall fit due to the small
statistics in each mass bin.
The resulting $p \bar{p} K^+ K^-$ invariant mass
distribution is shown in Fig.~\ref{m2p2k-fit}(a).

The cross section $\sigma_{\gamma \gamma \to p \bar{p} K^+ K^-}(W_{\gamma
\gamma})$ is calculated from
\begin{equation}\label{csfor}
\sigma_{\gamma \gamma \to p \bar{p} K^+ K^-}(W_{\gamma \gamma}) =
\frac{n^{\rm fit}}{\frac{dL_{\gamma \gamma}}{dW_{\gamma \gamma}}
\epsilon(W_{\gamma \gamma}) \Delta W_{\gamma \gamma}},
\end{equation}
where  $\frac{dL_{\gamma \gamma}}{dW_{\gamma \gamma}}$ is the
differential luminosity of the two-photon collision,
$\epsilon$ is the efficiency, $\Delta W_{\gamma \gamma}$ is the
bin width, and $n^{\rm fit}$ is the number of signal events in the $\Delta
W_{\gamma \gamma}$ bin.

The $\gamma \gamma \to p \bar{p} K^+ K^-$ cross section is shown in
Fig.~\ref{m2p2k-fit}(b), where the errors are
statistical only. No clear structure is seen in this
distribution and the largest value is around 40 pb, which is
lower than the rough estimate of 100 pb from Ref.~\cite{wu}.

\begin{figure}[htbp]
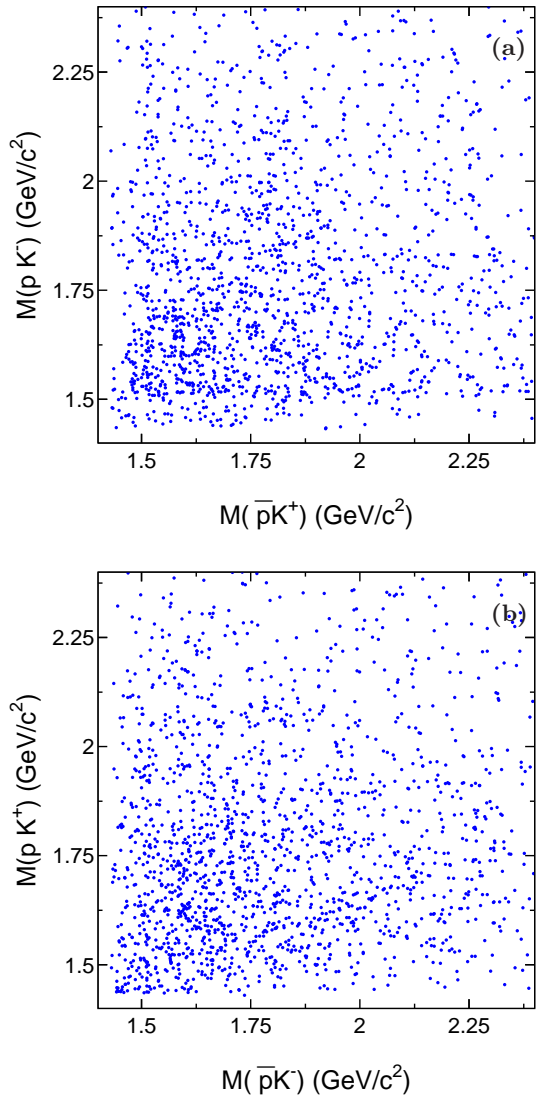

\psfig{file=fig/scatter3.epsi,width=7cm}
 \put(-20, 180){ \bf (a)}\vspace{0.5cm}
\psfig{file=fig/scatter4.epsi,width=7cm}
 \put(-20, 180){ \bf (b)}
\caption{The distributions of (a) $M(p K^-)$ versus $M(\bar{p} K^+)$
and (b) $M(p K^+)$ versus $M(\bar{p} K^-)$ for the selected $\gamma \gamma \to p \bar{p} K^+ K^-$
candidate events.} \label{scatter}
\end{figure}

To search for $Kp$ intermediate states, we require transverse
momentum balance for the $p \bar{p} K^+ K^-$ system
by imposing $|\sum {\vec P}_t^{\ast}|<0.17$~$\hbox{GeV}/c$,
which was optimized by maximizing the value of $S/\sqrt{S+B}$.
Here, $S$ is the number of fitted $\Lambda(1520)^0$
signal events and $B$ is the number of fitted background
events in the $\Lambda(1520)^0$ signal region.
Distributions of $M(p K^-)$ \textit{vs.} $M(\bar{p} K^+)$
and $M(p K^+)$ \textit{vs.} $M(\bar{p} K^-)$  for the selected
$p \bar{p} K^+ K^-$ events  are shown in Fig.~\ref{scatter}. Horizontal
and vertical bands can be seen in Fig.~\ref{scatter}(a) directions
at around 1.52 GeV/$c^2$, corresponding to
$\Lambda(1520)^0\to p K^-$ and $\bar{\Lambda}(1520)^0\to \bar{p} K^+$ decays
in the former plot. No signals are seen
for $\penz\to p K^-$ nor $\penpp \to p K^+$ in either scatter plot.

\begin{figure}[htbp]
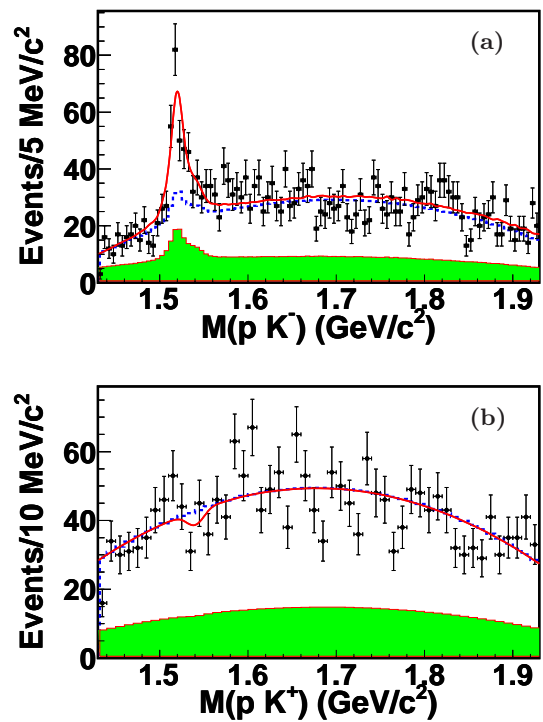

\psfig{file=fig/mpk-fit-total.epsi, angle=-90, width=7cm}
\put(-30, -15){ \bf (a)}\vspace{0.5cm}
\psfig{file=fig/mpkp-fit-total.epsi, angle=-90, width=7cm}
\put(-30, -15){ \bf (b)}
\caption{The fits to the (a) $p K^-$
and (b) $p K^+$ invariant mass distributions
for the (a) $\Lambda(1520)^0$, $\penz$
and (b) $\penpp$ candidates
in the whole $p \bar{p} K^+ K^-$ mass region.
The solid lines show the results of the simultaneous fits
described in the text, the
dotted curves show the total background estimates, and the shaded
histograms are the normalized $|\sum {\vec P}_t^{\ast}|$ sideband
contributions.}
\label{mpk-sim-fit}
\end{figure}

Figure~\ref{mpk-sim-fit} shows the $p K^-$ and $p K^+$ invariant mass distributions,
where for the $M(p K^+)$ distribution
the $p K^-$ mass is required to be outside $\Lambda(1520)^0$ signal region between 1.49 and 1.55 GeV/$c^2$.
To extract the signal and background yields in the
$|\sum {\vec P}_t^{\ast}|$ signal region and the corresponding sideband region, defined as
$0.6~\hbox{GeV}/c<|\sum {\vec P}_t^{\ast}|<1.0$ GeV$/c$,
unbinned maximum likelihood fits to the $p K^-$ invariant mass distributions are performed simultaneously.
The shapes of the $\Lambda(1520)^0$
and $\penz$ signals with mass resolutions of 6.5 MeV$/c^2$
are obtained from MC simulations
with energy-dependent widths and phase space factors
included. In the fit, a second-order Chebyshev polynomial
is used for the backgrounds in addition to the normalized
$|\sum {\vec P}_t^{\ast}|$ sideband contribution.
Similar fits with a $\Theta(1540)^{++}$ signal
are performed to the $p K^+$ invariant mass distributions.
Figure~\ref{mpk-sim-fit}(a) shows the fitted results to the
$M(p K^-)$ distribution with the $\Lambda(1520)^0$ and $\penz$
signal shapes included, while Fig.~\ref{mpk-sim-fit}(b)
shows the fitted results to the
$M(p K^+)$ distribution with the
$\penpp$ signal shape included only.
From the fits, the numbers of $\Lambda(1520)^0$, $\penz$ and $\penpp$
signal events are $288\pm48$, $22\pm34$ and $-16\pm34$, respectively.
The statistical significances of
the $\Lambda(1520)^0$ and $\penz$ are estimated to be $8.6\sigma$
and $1.4\sigma$, respectively,  by calculating
$\sqrt{-2\ln(\mathcal{L}_0/\mathcal{L}_{\rm max})}$,
where $\mathcal{L}_0$ and $\mathcal{L}_{\rm max}$ are the likelihoods of the fits
without and with the signal component, respectively.

\begin{figure}[htbp]
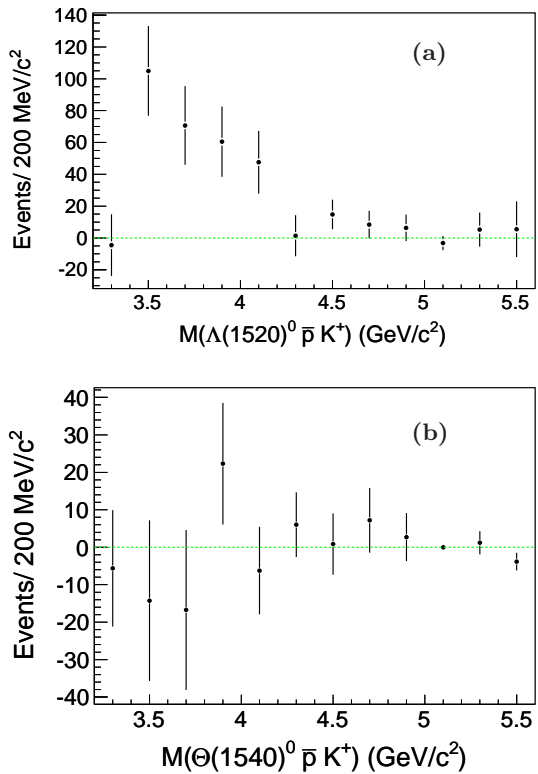

\psfig{file=fig/nlam-fit.epsi, angle=-90, width=7cm}
 \put(-50,-20){ \bf (a)}\vspace{0.5cm}
\psfig{file=fig/npen0-fit.epsi, angle=-90, width=7cm}
 \put(-50,-20){ \bf (b)}
\caption{  The (a) $\Lambda(1520)^0 \bar{p} K^+$ and (b) $\penz \bar{p} K^+$ invariant mass distributions obtained by
 fitting the $p K^-$ invariant mass distribution simultaneously to the $|\sum {\vec P}_t^{\ast}|$ signal region and sideband
 in each $p \bar{p} K^+ K^-$ mass bin.}
\label{lam-pen}
\end{figure}

\begin{figure}[htbp]
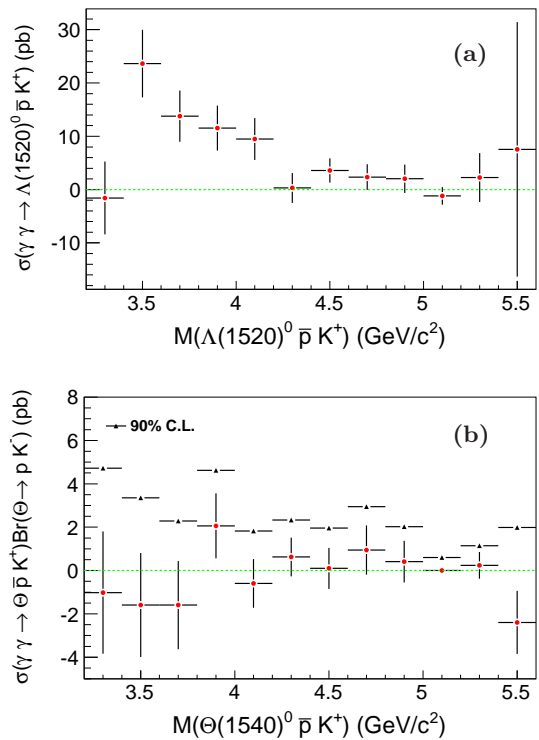

\psfig{file=fig/cs-lambda1520.epsi, angle=-90, width=7cm}
 \put(-35,-20){ \bf (a)}
\vspace{0.5cm}
\psfig{file=fig/cs-penta1.epsi, angle=-90, width=7cm}
 \put(-35,-20){ \bf (b)}
\caption{The measurements of (a) $\sigma(\gamma \gamma \to \Lambda(1520)^0 \bar{p} K^+$)
and (b) $\sigma(\gamma \gamma \to \penz \bar{p} K^+)$ $\BR(\penz \to p K^-)$
are shown as points with error bars. The error bars are statistical only.
The corresponding 90\% C.L. upper limits are shown with triangles.
}
\label{cs22}
\end{figure}

We obtain the yields of
$\Lambda(1520)^0 \bar{p} K^+$ and $\penz \bar{p} K^+$
in each $p \bar{p} K^+ K^-$ invariant mass
bin by performing similar simultaneous fits to $M(p K^-)$ distribution as was done above
for the entire $p \bar{p} K^+ K^-$ mass region.
The resulting $\Lambda(1520)^0 \bar{p} K^+$ and $\penz \bar{p} K^+$
invariant mass
distributions are shown in Figs.~\ref{lam-pen}(a) and (b).
The corresponding $\sigma(\gamma \gamma \to \Lambda(1520)^0 \bar{p} K^+$)
and $\sigma(\gamma \gamma \to \penz \bar{p} K^+)\BR(\penz \to p K^-)$
measurements are shown in Figs.~\ref{cs22}(a) and (b), where the error bars are
statistical only.
Since no $\penz$ signal is observed,
the upper limits  on the product $\sigma(\gamma \gamma \to \penz \bar{p} K^+)\BR(\penz \to p K^-)$
are shown in Fig.~\ref{cs22}(b) with triangles at the $90\%$ credibility level (C.L.),
where the upper limit on the signal yield ($N_{\rm up}$) in each $\penz \bar{p} K^+$ mass bin
is determined
by solving the equation $\int^{N_{\rm up}}_0\mathcal{L}(x)dx/\int^{+\infty}_0\mathcal{L}(x)dx$$=0.9$~\cite{cl},
where $x$ is the number of fitted signal events
and $\mathcal{L}(x)$ is the likelihood function in the fit to the data,
convolved with a Gaussian function whose width equals the total systematic uncertainty.


With a similar method, the number of
$\penpp \bar{p} K^-$ signal
events in each $p \bar{p} K^+ K^-$ invariant mass
bin is obtained from similar simultaneous fits to the $M(p K^+)$ distribution.
The resulting $\penpp \bar{p} K^-$
invariant mass distribution and the corresponding
$\sigma(\gamma \gamma \to \penpp \bar{p} K^-)\BR(\penpp \to p K^+)$ product
values, together with the upper limits at the 90\% C.L.,  are shown in Figs.~\ref{penpp}(a) and (b).

\begin{figure}[htbp]
\psfig{file=fig/npeta2-fit.epsi, angle=-90, width=7cm}
\put(-30,-15){ \bf (a)}\vspace{0.5cm}
\psfig{file=fig/cs-penta2.epsi, angle=-90, width=7cm}
\put(-30,-15){ \bf (b)}
\caption{(a) The $\penpp \bar{p} K^-$ invariant mass distribution obtained
by fitting the $p K^+$ invariant mass distribution simultaneously to the $|\sum {\vec P}_t^{\ast}|$ signal region and sideband
in each $p \bar{p} K^+ K^-$ mass bin and (b) the measurements of $\sigma(\gamma \gamma \to \penpp \bar{p} K^-)$$\BR(\penpp \to p K^+)$
shown as points with error bars. The error bars are statistical only. The corresponding 90\% C.L. upper limits are shown with triangles. }
\label{penpp}
\end{figure}

\begin{figure}[htbp]
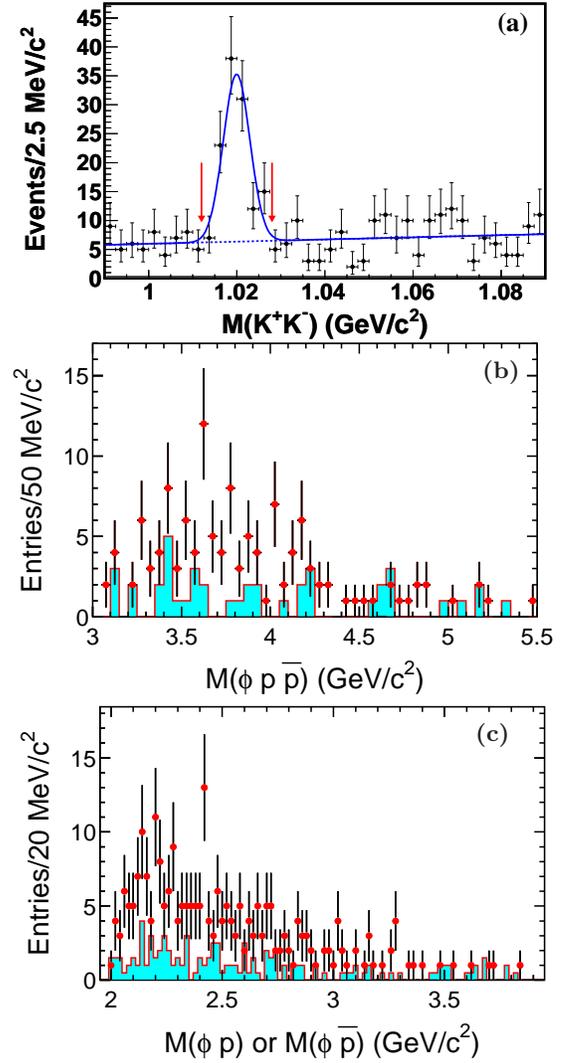

\psfig{file=fig/mkk-overall-fit.epsi,  angle=-90, width=7cm}
\psfig{file=fig/mphipp.epsi, height=4.7cm}
\put(-30,120){ \bf (b)} \vspace{0.1cm}
\psfig{file=fig/mphip-sum.epsi, height=4.7cm}
\put(-30,120){ \bf (c)}
\caption{ The (a) $K^+K^-$, (b)  $\phi p \bar{p}$, and (c) the sum of the $\phi p$ and $\phi \bar{p}$ invariant mass spectra.
The arrows in (a) indicate the  $\phi$ signal region.
The shaded histograms in (b) and (c) are from the normalized $\phi$-sideband events.
}
\label{mphip}
\end{figure}

Figure~\ref{mphip}(a) shows the $K^+K^-$ invariant mass distribution
of the selected candidate events, where a clear $\phi$
signal is observed.
An unbinned extended maximum likelihood fit is applied with a Gaussian resolution function with free parameters as
the $\phi$ signal shape and a first-order polynomial as the background shape.
From the fit, we obtain $88\pm12$ $\phi$ signal events. The $\phi$ signal region
is defined as the $\pm 8$ MeV/$c^2$ interval around the $\phi$ nominal mass~\cite{PDG},
as indicated by the arrows in Fig.~\ref{mphip}(a), and $\phi$ sidebands
are defined as $1.002~\hbox{GeV}/c^2<M(K^+K^-)<1.010~\hbox{GeV}/c^2$ or $1.030~\hbox{GeV}/c^2<M(K^+K^-)<1.038~\hbox{GeV}/c^2$.
The $\phi p \bar{p}$ invariant mass distribution within the $\phi$
signal region is shown in Fig.~\ref{mphip}(b), where the shaded histogram is
from the normalized $\phi$-sideband events. There are no evident structures.
The sum of the $\phi p$ and $\phi \bar{p}$ invariant mass distributions
is shown in Fig.~\ref{mphip}(c). No
significant evidence of  an $s\bar{s}$ partner of the pentaquark states
$P_c(4380)$ and $P_c(4450)$~\cite{lhcb} is observed.

Systematic error sources and their contributions to the
cross section measurements are summarized in Table~\ref{errtable}. The particle identification
uncertainties are 1.4\% for each kaon and 2.4\% (2.0\%) for each
proton (anti-proton).
A momentum-weighted systematic error in tracking efficiency
of about 0.4\% is
taken for each track.
The statistical error in the MC samples is about 0.5\%. The accuracy
of the two-photon luminosity function calculated with the {\sc
treps} generator is estimated to be about 5\%, including the error
from neglecting radiative corrections (2\%), the uncertainty from
the form factor effect (2\%)~\cite{treps}, and the uncertainty in the total
integrated luminosity (1.4\%).
The trigger efficiency
for four-charged-track events is rather high because of the
redundancy of the Belle first-level multi-track trigger. According
to the MC simulation,
the signal trigger efficiency increases with the $p \bar{p} K^+ K^-$ mass.
The uncertainty of
the trigger simulation is less than 4\%~\cite{epjc}.
The preselection efficiency for the final states has little dependence
on the  $p \bar{p} K^+ K^-$ invariant mass, with an uncertainty smaller
than 2.5\%. From Ref.~\cite{PDG}, the uncertainty in the
world average value for $\BR(\Lambda(1520)^0\to p K^-)$ is 2.3\%.
The uncertainty in the fitted yield for the signal is estimated by
varying the background shape and fit range,
which is 8.1\% for $p \bar{p} K^+ K^-$, 9.8\% for $\Lambda(1520)^0 \bar{p} K^+$,
40\% for $\penz \bar{p} K^+$, and 21\% for $\penpp \bar{p} K^-$.
For the $\Lambda(1520)^0 \bar{p} K^+$, $\penz \bar{p} K^+$, and
$\penpp \bar{p} K^-$ modes, we
estimate the systematic errors associated with the $\Lambda(1520)^0$
and $\Theta(1540)$ resonance
parameters by changing the values of the masses and widths of the
resonances by $\pm 1\sigma$.  The resulting differences of
5.7\%, 7.4\% and 5.5\% in the fitted results are taken as systematic errors.
The uncertainty on the mass resolution is estimated
by changing the MC signal resolution by $\pm 10\%$,
which is 1.1\% for $\Lambda(1520)^0 \bar{p} K^+$,
2.5\% for $\penz \bar{p} K^+$, and 4.0\%
for $\penpp \bar{p} K^-$.
Assuming that
all of these systematic error sources are independent, we add them
in quadrature to obtain the total systematic errors of
13\%, 16\%, 42\% and 25\% for $p \bar{p} K^+ K^-$, $\Lambda(1520)^0 \bar{p} K^+$,
$\penz \bar{p} K^+$, and $\penpp \bar{p} K^-$, respectively.

\begin{table}[htbp]
\caption{Relative systematic errors (\%) on the cross section measurements for
$\gamma \gamma \to p \bar{p} K^+ K^-$, $\Lambda(1520)^0 \bar{p} K^+$,
$\penz \bar{p} K^+$ and $\penpp \bar{p} K^-$, respectively.} \label{errtable}
\begin{tabular}{c | c | c | c | c }
\hline Source & $p \bar{p} K^+ K^-$ &  $\Lambda(1520)^0$ &  $\Theta^0$ & $\Theta^{++}$  \\\hline
 Part. identification &  7.2  & 7.2  & 7.2 & 7.2  \\
 Tracking & 1.6  & 1.6 & 1.6 & 1.6   \\
  MC statistics &  0.5 & 0.5  & 0.5 & 0.5  \\
 Lum. function & 5.0 & 5.0 & 5.0 & 5.0 \\
 Trigger efficiency & 4.0 & 4.0 & 4.0 & 4.0  \\
 Preselection efficiency & 2.5  & 2.5 & 2.5 & 2.5  \\
  Branching fractions & -- & 2.3  & --  & --  \\
 Fit uncertainty & 8.1 & 9.8 & 40 & 21  \\
 Res. parameters &-- & 5.7 & 7.4 & 5.5 \\
 Signal resolution & -- & 1.1 & 2.5 & 4.0  \\
 \hline
 Sum & 13 & 16 & 42 & 25 \\
 \hline
\end{tabular}
\end{table}

For a $p \bar{p} K^+ K^-$ invariant mass above 3.1~GeV/$c^2$, we measure the
production rate of charmonium states.
In measuring the production rates, $|\sum
{\vec P}_t^{\ast}|$ is required to be less than $0.17~\hbox{GeV}/c$
in order to reduce backgrounds from non-two-photon processes and
two-photon processes with extra particles other than the signal
final state.

\begin{figure}[htbp]
\psfig{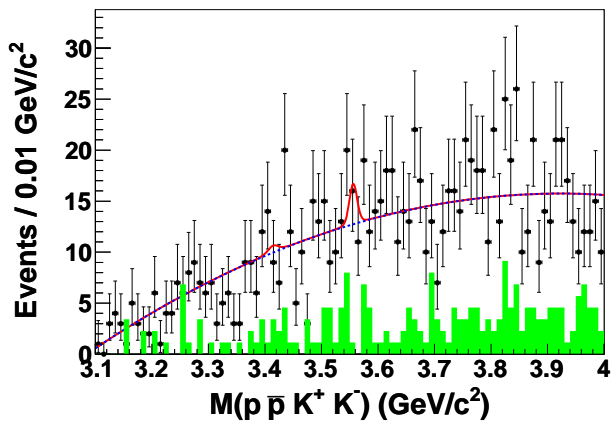}  \caption{
The invariant mass distribution of $p \bar{p} K^+ K^-$ in the charmonium mass region with the requirement $|\sum {\vec
P}_t^{\ast}|<0.17~\hbox{GeV}/c$. The shaded histogram is the normalized
$|\sum {\vec P}_t^{\ast}|$ sideband contribution. The points
with error bars are data, the dashed line is the fitted total background and the solid curve is the best fit.}
\label{charmonium}
\end{figure}

Figure~\ref{charmonium} shows the $p \bar{p} K^+ K^-$ invariant mass
distribution. No clear $\chi_{c0}$ or $\chi_{c2}$
signals are seen. The mass spectrum
is fitted with two incoherent Breit-Wigner functions convolved
with a corresponding Gaussian resolution function as the
$\chi_{c0}$ and $\chi_{c2}$ signal shapes, and a
second-order Chebyshev polynomial as the background shape.
The fit result is shown in Fig.~\ref{charmonium} as a
solid curve, where  the dashed line is the fitted  background and
the shaded histogram is the normalized
$|\sum {\vec P}_t^{\ast}|$ sideband contribution.
The statistical signal significances are
0.2$\sigma$ and 0.8$\sigma$
for the $\chi_{c0}$ and $\chi_{c2}$, respectively.

Since no significant signals are observed,
the 90\% C.L. upper limits on the $\chi_{c0}$ and $\chi_{c2}$ yields
are determined to be 16.4 and 18.0, respectively. In these calculations,
we assume there is no interference between the charmonium and the
continuum amplitudes.
A systematic error estimate similar to that for the cross sections
results in total systematic errors of 29\% and 13\% for
$\Gamma_{\gamma \gamma}(R) \BR(R \to p \bar{p} K^+ K^-)$
for $R=\chi_{c0}$ and $\chi_{c2}$,
respectively.

The product of the two-photon decay width and branching fraction is
obtained from the relation $\Gamma_{\gamma \gamma}(R)\BR(R\to
\hbox{final~state})=N/[(2J+1)\epsilon\, {\cal K} \lum_{\rm int}]$,
where $N$ is the number of observed events, $\epsilon$ is the
efficiency, $J$ is the spin of the resonance, and $\lum_{\rm int}$
is the integrated luminosity. The factor ${\cal K}$ is
calculated from the two-photon luminosity function ${\cal
L}_{\gamma\gamma}(M_R)$ for a resonance with mass $M_R$ using the
relation ${\cal K}=4\pi^2{\cal L}_{\gamma\gamma}(M_R)/M_R^2$, which
is valid when the resonance width is small compared to its mass.
The ${\cal K}$ factor is calculated to be
$1.15~\hbox{fb}/\hbox{eV}$ and $0.95~\hbox{fb}/\hbox{eV}$
for the $\chi_{c0}$ and  $\chi_{c2}$, respectively, using {\sc
treps}~\cite{treps}. The efficiencies are 2.77\% and 3.97\% for the $\chi_{c0}$ and $\chi_{c2}$, respectively.
From the above results, we obtain
upper limits of 0.53 eV ($J=0$) and 0.10 eV ($J=2$) for
$\Gamma_{\gamma\gamma}(\chi_{cJ})\BR(\chi_{cJ}\to p \bar{p} K^+ K^-)$.


In summary, we observe the process $\gamma \gamma \to
p \bar{p} K^+ K^-$ and search for the first time for possible
 exotic baryons $\penz$ and $\penpp$
decaying to $p K^-$ and $p K^+$ in the two-photon process
$\gamma \gamma \to p \bar{p} K^+ K^-$.
Clear $\gamma \gamma \to p \bar{p} K^+ K^-$ signals
are observed. While the $\Lambda(1520)^0$ signals in
$pK^-$ invariant mass spectrum are also observed,
no evidence for any exotic baryon is seen
in the $p K^-$ or $pK^+$ invariant mass spectrum.
For all of the above-mentioned processes, the
cross sections are measured for the first time.
The cross sections for $\gamma \gamma \to  p \bar{p} K^+ K^-$ are
lower by a factor 2.5 or more than the theoretical prediction of 0.1 nb in Ref.~\cite{wu}.
At the same time,  no clear $\chi_{c0}$ or $\chi_{c2}$ signal is seen
in the $p \bar{p} K^+ K^-$ invariant mass spectrum,
and 90\% C.L. upper limits on the products of
the two-photon decay width and branching fraction of the $\chi_{c0}$ and $\chi_{c2}$ to $p \bar{p} K^+ K^-$
are established.


We thank the KEKB group for the excellent operation of the
accelerator; the KEK cryogenics group for the efficient
operation of the solenoid; and the KEK computer group,
the National Institute of Informatics, and the
PNNL/EMSL computing group for valuable computing
and SINET4 network support.  We acknowledge support from
the Ministry of Education, Culture, Sports, Science, and
Technology (MEXT) of Japan, the Japan Society for the
Promotion of Science (JSPS), and the Tau-Lepton Physics
Research Center of Nagoya University;
the Australian Research Council;
Austrian Science Fund under Grant No.~P 22742-N16 and P 26794-N20;
the National Natural Science Foundation of China under Contracts
No.~10575109, No.~10775142, No.~10875115, No.~11175187, No.~11475187
and No.~11575017;
the Chinese Academy of Science Center for Excellence in Particle Physics;
the Ministry of Education, Youth and Sports of the Czech
Republic under Contract No.~LG14034;
the Carl Zeiss Foundation, the Deutsche Forschungsgemeinschaft, the
Excellence Cluster Universe, and the VolkswagenStiftung;
the Department of Science and Technology of India;
the Istituto Nazionale di Fisica Nucleare of Italy;
the WCU program of the Ministry of Education, National Research Foundation (NRF)
of Korea Grants No.~2011-0029457,  No.~2012-0008143,
No.~2012R1A1A2008330, No.~2013R1A1A3007772, No.~2014R1A2A2A01005286,
No.~2014R1A2A2A01002734, No.~2015R1A2A2A01003280 , No. 2015H1A2A1033649;
the Basic Research Lab program under NRF Grant No.~KRF-2011-0020333,
Center for Korean J-PARC Users, No.~NRF-2013K1A3A7A06056592;
the Brain Korea 21-Plus program and Radiation Science Research Institute;
the Polish Ministry of Science and Higher Education and
the National Science Center;
the Ministry of Education and Science of the Russian Federation and
the Russian Foundation for Basic Research;
the Slovenian Research Agency;
Ikerbasque, Basque Foundation for Science and
the Euskal Herriko Unibertsitatea (UPV/EHU) under program UFI 11/55 (Spain);
the Swiss National Science Foundation;
the Ministry of Education and the Ministry of Science and Technology of Taiwan;
and the U.S.\ Department of Energy and the National Science Foundation.
This work is supported by a Grant-in-Aid from MEXT for
Science Research in a Priority Area (``New Development of
Flavor Physics'') and from JSPS for Creative Scientific
Research (``Evolution of Tau-lepton Physics'').


\begin{thebibliography}{**}
\bibitem{gell} M. Gell-Mann, Phys. Lett. {\bf 8}, 214 (1964).

\bibitem{zc4430} S.-K. Choi {\it et al.} (Belle Collaboration),
Phys. Rev. Lett. {\bf 100}, 142001 (2008).

\bibitem{belley_new} M. Ablikim  {\it et al.} (BESIII Collaboration),
Phys. Rev. Lett. {\bf 110}, 252001 (2013); Z. Q. Liu  {\it et al.} (Belle Collaboration),
Phys. Rev. Lett. {\bf 110}, 252002 (2013).

\bibitem{zc4020} M. Ablikim  {\it et al.} (BESIII Collaboration),
Phys. Rev. Lett. {\bf 111}, 242001 (2013).

\bibitem{zc4200} K. Chilikin {\it et al.} (Belle Collaboration),
Phys. Rev. D {\bf 90}, 112009 (2014).


\bibitem{roman} R. Mizuk {\it et al.} (Belle Collaboration),
Phys. Rev. D {\bf 78}, 072004 (2008).

\bibitem{QWG} N. Brambilla {\it et al.}, \Journal\EPJC{71}{1534}{2011};
 \Journal\EPJC{74}{2981}{2014}.

\bibitem{lhcb} R. Aaij {\it et al.} (LHCb Collaboration),
Phys. Rev. Lett. {\bf 115}, 072001 (2015).

\bibitem{doublet} R. F. Lebed, Phys. Lett. B {\bf 749}, 454 (2015);
 L. Maiani, A. D. Polosa and V. Riquer, Phys. Lett. B {\bf 749}, 289 (2015);
 V. V. Anisovich, M. A. Matveev, J. Nyiri, A. V. Sarantsev and A. N. Semenova, arXiv:1507.07652;
 R. Ghosh, A. Bhattacharya and B. Chakrabarti, arXiv:1508.00356; V.V. Anisovich, M.A. Matveev, J. Nyiri, A.V. Sarantsev
 and A.N. Semenova, arXiv:1509.04898.


\bibitem{mole} H. X. Chen, L. S. Geng, W. H. Liang, E. Oset, E. Wang
and J. J. Xie, arXiv:1510.01803;  R. Chen, X. Liu, X. Q. Li and S. L. Zhu£¬
Phys. Rev. Lett. {\bf 115}, 132002 (2015);  L. Roca, J. Nieves and E. Oset,
arXiv:1507.04249;  J. He, arXiv:1507.05200;  U. G. Meiner and J. A. Oller, arXiv:1507.07478.


\bibitem{leps} T. Nakano {\it et al.} (LEPS Collaboration),
Phys. Rev. Lett. {\bf 91}, 012002 (2003).

\bibitem{mao} T. B. Liu, Y. J. Mao and B. Q. Ma, Int. J. Mod. Phys. A {\bf 29},
1430020 (2014).

\bibitem{wu} S. Armstrong, B. Mellado and S. L. Wu, J. Phys. G {\bf 30}, 1801 (2004).

\bibitem{PDG}  K. A. Olive {\it et al.}  (Particle Data Group), Chin. Phys. C {\bf 38}, 090001 (2014) and 2015 update.


\bibitem{Belle} A.~Abashian {\em et al.} (Belle Collaboration),
 Nucl. Instrum. Methods Phys. Res., Sect. A {\bf 479}, 117 (2002);
 also, see detector section in J. Brodzicka {\em et al.}, Prog. Theor. Exp. Phys. (2012) 04D001.


\bibitem{KEKB} S. Kurokawa and E. Kikutani, Nucl. Instrum. Methods
Phys. Res., Sect. A {\bf 499}, 1 (2003), and other papers
included in this volume; T. Abe {\em et al.}, Prog. Theor. Exp. Phys. (2013) 03A001
and following articles up to 03A011.

\bibitem{treps} S.~Uehara, KEK Report 96-11 (1996).
In the generator, the form factor is assumed to be $1/(1+Q^2/W^2)$,
where $Q^2$ is a 4-momentum transfer of an electron and
represents the virtuality of the photon.
[http://lss.fnal.gov/archive/other1/kek-report-96-11.pdf].

\bibitem{berger} C.~Berger and W.~Wagner, Phys. Rept. {\bf 146}, 1 (1987).

\bibitem{charge} Charge-conjugate
decays are implicitly assumed throughout the paper.


\bibitem{pid} E.~Nakano,
Nucl. Instrum. Methods Phys. Res., Sect. A {\bf 494}, 402 (2002).

\bibitem{EID} K. Hanagaki {\em et al.},
Nucl. Instrum. Methods Phys. Res., Sect. A {\bf 485}, 490 (2002).

\bibitem{cl} In common high energy physics usage, this
Bayesian interval has been reported as ``confidence interval,'' which
is a frequentist-statistics term.


\bibitem{epjc} S.~Uehara {\em et al.} (Belle Collaboration),
Eur. Phys. J. C {\bf 53}, 1 (2008).



\end{thebibliography}
\end{document}